\documentclass[aps,prd,onecolumn,groupedaddress,nofootinbib,preprintnumbers,superscriptaddress]{revtex4-2}  
\usepackage{graphicx}
\usepackage{amsmath}
\usepackage{amsfonts}
\usepackage[colorlinks=true,citecolor=red,linkcolor=blue,breaklinks=true]{hyperref}
\usepackage[figure, table]{hypcap}
\usepackage{amssymb}
\usepackage{xcolor}
\usepackage{subfigure}
\usepackage{setspace}
\usepackage{footnote}
\usepackage{multirow}
\usepackage[normalem]{ulem}
\usepackage[utf8]{inputenc}
\usepackage{mathrsfs}
\usepackage{longtable}
\usepackage{enumitem}

\newcommand{\CNB}{{\rm C}\nu {\rm B}}

\DeclareMathAlphabet{\mathpzc}{OT1}{pzc}{m}{it}

\newcommand{\MeV}{{\rm MeV}}

\allowdisplaybreaks

\definecolor{palatd}{RGB}{104, 36, 109}
\definecolor{palatb}{RGB}{0, 56, 168}
\definecolor{palatr}{rgb}{0.745,0.118,0.176}
\newcommand\myshade{80}
\colorlet{mylinkcolor}{palatr}
\colorlet{mycitecolor}{palatb}
\colorlet{myurlcolor}{palatd}

\hypersetup{
  linkcolor  = mylinkcolor!\myshade!black,
  citecolor  = mycitecolor!\myshade!black,
  urlcolor   = myurlcolor!\myshade!black,
  colorlinks = true
}

\def\MeV{{\rm MeV}}

\begin{document}

\preprint{IPPP/24/28}

\title{
Effects of Neutrino-Ultralight Dark Matter Interaction on 
\\
the Cosmic Neutrino Background
}

\author{Pablo Mart\'{i}nez-Mirav\'e}
\email{pablo.mirave@nbi.ku.dk}
\affiliation{Niels Bohr International Academy and DARK, Niels Bohr Institute, University of Copenhagen, \\
Blegdamsvej 17, 2100, Copenhagen, Denmark} 

\author{Yuber F. Perez-Gonzalez}
\email{yuber.f.perez-gonzalez@durham.ac.uk}
\affiliation{Institute for Particle Physics Phenomenology, Durham University, South Road DH13EL, Durham, United Kingdom}

\author{Manibrata Sen}
\email{manibrata@mpi-hd.mpg.de}
\affiliation{Max-Planck-Institut f\"ur Kernphysik, Saupfercheckweg 1, 69117 Heidelberg, Germany}

\begin{abstract}
Ultralight dark matter interacting with sterile neutrinos would modify the evolution and properties of the cosmic neutrino background through active-sterile neutrino mixing.  We investigate how such an interaction would induce a redshift dependence in neutrino masses.  We highlight that cosmological constraints on the sum of neutrino masses would require reinterpretation due to the effective mass generated by neutrino-dark matter interactions. Furthermore, we present an example where such interactions can alter the mass ordering of neutrinos in the early Universe, compared to what we expect today. We also address the expected changes in the event rates in a PTOLEMY-like experiment, which aims to detect the cosmic neutrino background via neutrino capture and discuss projected constraints.
\end{abstract}

\maketitle
\section{Introduction}
\label{sec:intro}

The origin of neutrino masses and the nature of dark matter constitute, at present, the two main open questions in our understanding of the Universe. On the one hand, neutrino oscillation data is consistent with the existence of three active neutrinos, being --- at least two of them -- massive particles at present times~\cite{deSalas:2020pgw,Esteban:2020cvm,Capozzi:2021fjo}. Additional sterile neutrino species could exist and explain several experimental anomalies~\cite{LSND:2001aii,MiniBooNE:2018esg, GALLEX:1997lja,SAGE:1998fvr,Barinov:2021asz, Giunti:2021kab}. However, regions of the parameter space determined by their mass and mixing with the three active neutrinos are severely constrained from neutrino disappearance experiments as well as cosmological observables~\cite{Diaz:2019fwt,Abazajian:2017tcc,Gariazzo:2019gyi,Hagstotz:2020ukm}. On the other end, a plethora of dark matter candidates have been hypothesised and their signatures are the object of many direct and indirect experimental searches. Nonetheless, a positive signal remains elusive.

In this context, we explore the phenomenology of scalar dark matter candidates with masses $\sim 10^{-22} - 10^{-10}\,$eV, which are often referred to as ultralight dark matter, fuzzy dark matter or wave-like dark matter~\cite{Hui:2016ltb, Ferreira:2020fam,Hui:2021tkt}. Such candidates were initially proposed as an explanation to the small-scale cosmological puzzles such as the too-big-to-fail problem, the missing-satellites riddle or the cup-vs-core puzzle~\cite{Bullock:2017xww}. In the light of recent data, the initial motivation has  -- at least partially -- vanished~\cite{Fairbairn:2022gar}. Nevertheless, ultralight dark matter exhibits a very rich phenomenology, testable in a wide variety of experiments - ranging from precision atomic experiments to gravitational wave observations. From a theoretical point of view, ultralight scalars or pseudoscalars can arise in various minimal extensions of the Standard Model (SM), for instance from spontaneous lepton number violation\cite{Reig:2019sok}, in the Nelson-Barr solution to the CP-problem~\cite{Dine:2024bxv} or in the Peccei-Quinn mechanism~\cite{Peccei:1977hh}.

From a phenomenological point of view, coupling the ultralight dark matter (ULDM) to SM fermions can give rise to a variety of signatures, such as enhanced stellar cooling, additional contributions to electric and magnetic dipole moments, or lepton flavour violating decays, among others~\cite{Escribano:2020wua}. Recent studies have focused on the signatures of time-variations of neutrino masses, which result from neutrinos coupling to such ULDM scalar field~\cite{Berlin:2016woy,Capozzi:2017auw,Brdar:2017kbt,Farzan:2018pnk,Dev:2020kgz,Dev:2022bae, Long:2014zva, Huang:2021kam, Huang:2022wmz, Losada:2022uvr, Cordero:2022fwb, Dev:2022bae, Davoudiasl:2023uiq, Losada:2023zap, ChoeJo:2023ffp, Sen:2023uga, Gherghetta:2023myo,Lambiase:2023hpq, Lin:2023xyk, Arguelles:2024cjj}. However, such a scenario can also modify neutrino free-streaming or lead to redshift-dependent neutrino masses. These, in turn, could be in conflict with observations of the cosmic microwave background and large-scale structures~\cite{Craig:2024tky}. A simple way to evade this constraint is to couple the ULDM to sterile neutrinos, which are SM singlets (see e.g. ~\cite{Huang:2022wmz,He:2023neh}).
Note that, due to the mixing between active and sterile neutrinos, this interaction could manifest as an effective coupling between ultralight dark matter and active neutrinos, which could leave an imprint in oscillation experiments, beta-decay measurements and neutrinoless double-beta decay searches. This portal would simultaneously avoid the direct coupling between dark matter and charged fermions, which has been tightly constrained from the non-observation of time variations in the electron mass, see e.g.~\cite{Scoccola:2008yf}.

In this article, we test interactions between ULDM and active neutrinos using the sea of relic neutrinos from the Big Bang that permeates the Universe. This sea of relic neutrinos -- aptly known as the cosmic neutrino background ($\CNB$) -- is a crucial prediction of the $\Lambda$CDM model of cosmology, and positive detection of this background can be used to test fundamental properties of neutrinos~\cite{Weinberg:1962zza,Ringwald:2004np,Mertsch:2019qjv,Brdar:2022kpu,Arteaga:2017zxg,Akita:2021hqn, Alvey:2021xmq, Das:2022xsz, Banerjee:2023lrk,Long:2014zva, Roulet:2018fyh,Perez-Gonzalez:2023llw,Ciscar-Monsalvatje:2024tvm,Franklin:2024enc}. A number of ideas has been proposed for the detection of this neutrino background~\cite{Weinberg:1962zza,Stodolsky:1974aq,Shvartsman:1982sn,Akhmedov:2019oxm, Chao:2021ahl, Shergold:2021evs,Bauer:2021uyj,Brdar:2022wuv}, however, the most feasible one till date is that of neutrino capture on a beta-decaying nuclei like tritium, as put forward by Weinberg~\cite{Weinberg:1962zza} and currently the major focus of the PTOLEMY collaboration~\cite{PTOLEMY:2018jst}. 
From neutrino oscillation experiments, it is expected that at least two generations of neutrinos composing this $\CNB$ are non-relativistic at present times. However, this picture might change if the neutrino masses acquired an additional contribution from their interaction with the ULDM. Such an effective neutrino mass, sourced from ULDM, alters the evolution of the $\CNB$~in the late Universe. 
Not only that, the effects of these interactions can show up in a PTOLEMY-like experiment, aimed at detecting the $\CNB$. Here, we propose to use the $\CNB$~as a laboratory to test ULDM interactions with neutrinos and set novel constraints on the parameter space defined by ULDM mass and coupling.

The paper is organised as follows.
In Section~\ref{sec:theo}, we present the framework in which we discuss dark matter-neutrino coupling and the general phenomenological implications. In Section~\ref{sec:cnub}, we describe how this coupling would induce neutrino masses that grow with redshift. Indirect probes of neutrino masses can constrain their redshift evolution. We also comment on the possibility of altering the hierarchy of neutrino masses at early times. In Section~\ref{sec:PTOLEMY}, we show the observable imprint of this scenario in a PTOLEMY-like experiment, based on the process of neutrino capture. Finally, in Section~\ref{sec:conc}, we outline the main conclusions and discuss future work along these lines.
We use the natural unit system where $\hbar = c = k_{\rm B} = 1$, and define the Planck mass to be $m_{\rm PL}=1/\sqrt{8 \pi G}$, with $G$ being the gravitational constant, throughout this manuscript.

\section{THEORETICAL FRAMEWORK}
\label{sec:theo}

\begin{figure}
    \centering
    \includegraphics[width = \textwidth]{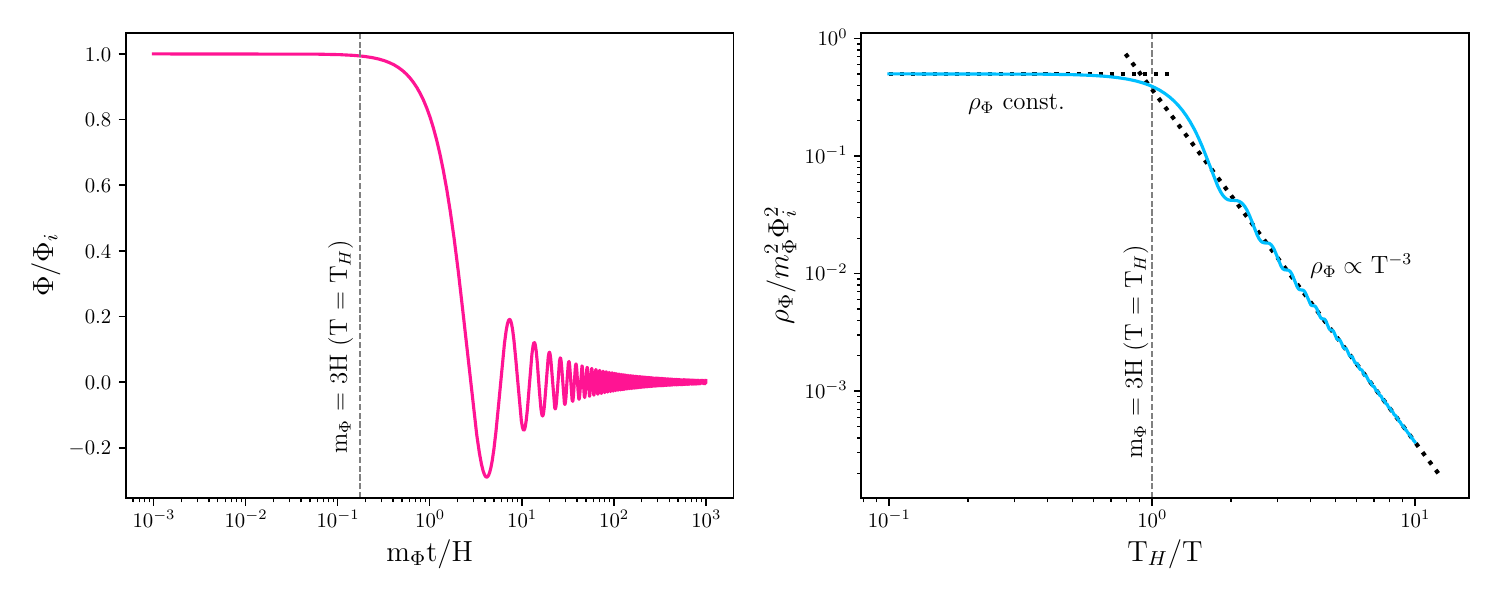}
    \caption{Left panel: Evolution of the ultralight dark matter field in the misalignment mechanism, according to its initial value $\Phi_i$. Radiation is assumed to dominate the background evolution. The vertical dashed line indicates when the mass and Hubble-friction terms in Equation \ref{eq:evolphi} become comparable. Right panel: Evolution of the energy density of the ULDM field in a radiation-dominated background, given an initial value $\rho_i = \frac{1}{2}m_\Phi^2 \Phi^2$. The dotted lines indicate the regimes for which the energy density is constant and when it scales as a function of the temperature, namely $T^3$.}
    \label{fig:phi-evol}
\end{figure}

In the early Universe, a scalar field $\Phi$ with mass $m_\Phi$ evolves according to the following equation of motion,
\begin{align}
    \ddot{\Phi} + 3H(t)\dot{\Phi} + m^2_\Phi\Phi = 0 \, ,
    \label{eq:evolphi}
\end{align}
where the dot represents the derivative with respect to proper time and $H$ is the Hubble rate. 
Then, the relic density of an ultralight dark matter $\Phi$ can be determined through the misalignment mechanism~\cite{Preskill:1982cy,Abbott:1982af,Dine:1982ah}. This mechanism involves the initial displacement of $\Phi$ from the potential's minimum, where the field remains static due to Hubble friction -- i.e. the second term in Equation~\ref{eq:evolphi}. At this point, the DM density is frozen to its initial value. At later times, when the temperature drops below a critical value $T_H$ -- defined by the equation $m_\Phi=3H(T_H)$ -- $\Phi$ begins oscillating. As the temperature decreases further ($T<T_H$), the DM transitions to a non-relativistic state, causing its energy density to vary as $T^3$~\cite{Dev:2022bae,Sen:2023uga}. This behavior is roughly described by
\begin{equation}
\rho_\Phi(T)=\rho_\Phi (T_0) \left(\frac{g(T_0)}{g(T)}\right)\left(\frac{{\rm min}(T,T_H)}{T_0}\right)^3\, ,
\end{equation}
with $\rho_\Phi (T_0)=10^{-5} \rho_{\Phi, \textrm{local}}$ the current DM density, scaled by the local DM overdensity factor, $g(T)$ denotes the entropy effective relativistic degrees of freedom, and $T_0=2.72\,$K represents the current temperature of the cosmic microwave background (CMB). Consequently, the energy density of DM is constant at early times until $m_\phi=3H$, after which it starts decreasing following the temperature evolution. Fig.\,\ref{fig:phi-evol} displays the evolution of the ultralight dark matter field and its energy density in a flat radiation-dominated Friedmann-Lemaître-Robertson-Walker background (see e.g.~\cite{OHare:2024nmr} for more details).

At present times, and due to the large occupation number, the ultralight scalar field is well-described as a classical field, 
\begin{align}
    \Phi(t, z = 0) \simeq \frac{\sqrt{2 \rho_{\Phi, \textrm{local}}}}{m_\Phi} \cos \left(m_\Phi t\right)\,,
\end{align}
being $\rho_{\Phi, \textrm{local}}$ the local dark matter density, and we have neglected a phase in the time modulation. An elaborate discussion on why this phase can be neglected can be found in~\cite{Sen:2023uga}.

Motivated by the feebly interacting nature of sterile neutrinos and dark matter, we assume an interaction between both species that acts as a portal between dark matter and the Standard Model due to active-sterile neutrino mixing.
In particular, let us consider a simplified case where a sterile neutrino, $N$, couples to an ultralight dark matter scalar field, $\Phi$, and mixes with only one family of leptons.
Such a scenario is described by the following Lagrangian,
\begin{align}
    - \mathcal{L} \supset\,  y_D \overline{l_L}\widetilde{h}N + \frac{1}{2}\left(m_N+ y_\Phi\Phi\right)\overline{N^c}N + \frac{1}{2}m_\Phi \Phi^2 + \frac{\kappa}{2}\left(\overline{l^c_L}\widetilde{H}^*\right)\left(\widetilde{H}^\dagger l_L\right) + \textrm{h.c.} \, .
\end{align}
Here,
$m_N$ is the mass of sterile neutrino, and $y_\Phi$ is the coupling between both species. In addition, $h$ denotes the Higgs field and $l_L$ denotes the Standard Model lepton doublet $l_L \equiv (\nu, e)_L$. The Dirac mass, $m_D = y_D v/\sqrt{2}$, and the active-neutrino Majorana mass, $m_\nu = \kappa v^2/2$, results from the Higgs field getting a vacuum expectation value, $v$. 
Note that the dimension-5 operator ensures that the sterile neutrino mass and mixing angle can be chosen independently in our 1+1 scenario. Extensions to more generations will be considered in future work.

The neutrino mass matrix in this simplified framework with one active and one sterile neutrino (1+1) is diagonalised by the mixing matrix, $U$, parameterised by the angle $\theta_{14}$. We denote the corresponding mass eigenstates as $m_1$ and $m_4$. Once we include a dark-matter--sterile neutrino coupling, the mass matrix is written in the new basis as
\begin{align}
    \mathcal{M}_{\nu+ \textrm{DM}} = U^\dagger \begin{pmatrix}
        m_1 & 0 \\ 0 & m_4 
    \end{pmatrix}U + \begin{pmatrix}
        0 & 0 \\ 0& y_\Phi\Phi
    \end{pmatrix}\, = \widetilde{U}^\dagger \begin{pmatrix}
        \widetilde{m}_1 & 0 \\ 0 & \widetilde{m}_4
    \end{pmatrix}\widetilde{U}\, ,
\end{align}
from which stems the definition of the effective masses, $\widetilde{m}_1$ and $\widetilde{m}_4$, and the effective mixing matrix, $\widetilde{U}$, parameterised in terms of the mixing angle $\widetilde{\theta}_{14}$.
As a consequence of the mixing, the effective mass eigenstates acquire a time dependence, namely
\begin{align}
    \widetilde{m}_{1,4}= \frac{m_1 + m_4 + y_\Phi\Phi}{2}\pm\frac{1}{2}\sqrt{(m_4-m_1)^2 + (y_\Phi\Phi)^2 +2(m_4-m_1)y_\Phi\Phi\cos 2\theta_{14}}\, 
    \label{eq:NuMass}.
\end{align}
Additionally, the effective mixing also becomes time-dependent,
\begin{align}
    \tan2\widetilde{\theta}_{14} = \frac{(m_4 -m_1)\sin2\theta_{14}}{(m_4-m_1)\cos2\theta_{14} + y_\Phi\Phi}\, .
\end{align}
Note that due to the self-interactions, the effective sterile neutrino mass and mixing can be suppressed even though their ``vacuum'' values are large. This can have important consequences on the sterile neutrino interpretation of the short baseline anomalies, as explored in~\cite{Huang:2022wmz}. 
In the next section, we discuss some of the implications of such an effective neutrino mass and mixing.

\section{Evolution of the effective neutrino mass and relevance to the $\CNB$}
\label{sec:cnub}
Due to the redshift evolution of the dark matter field, the effective masses $\widetilde{m}_1$ and $\widetilde{m}_4$, and the effective mixing, $\widetilde{\theta}_{14}$, would also change along with the evolution of the Universe. This means that active neutrinos would get an effective mass, as a consequence of mixing with sterile neutrinos. On the one hand, for a suitable choice of parameters, in the present-day scenario, $y_\Phi\Phi \ll m_4$ can be reached. It can be shown that this case would manifest as a time variation in neutrino observables at terrestrial experiments. 
In the early Universe, due to redshift variation, the condition could evolve to be $y_\Phi\Phi \gg m_4$. In this limit, it has been shown that the induced neutrino mass in the early Universe is small as long as $m_4\,\sin^2\theta_{14}$ is small~\cite{Huang:2022wmz}. This allows us to bypass the limits arising on the sum of neutrino masses from cosmology~\cite{Huang:2022wmz}. This particular limit can not only lead to large neutrino mass cosmology~\cite{Alvey:2021xmq}, but also give rise to scenarios where the neutrino mass ordering in the early Universe can be different from what it is today and modify the interpretation of cosmological probes of neutrino masses.

\begin{figure}[!t]
    \includegraphics[width=0.6\textwidth]{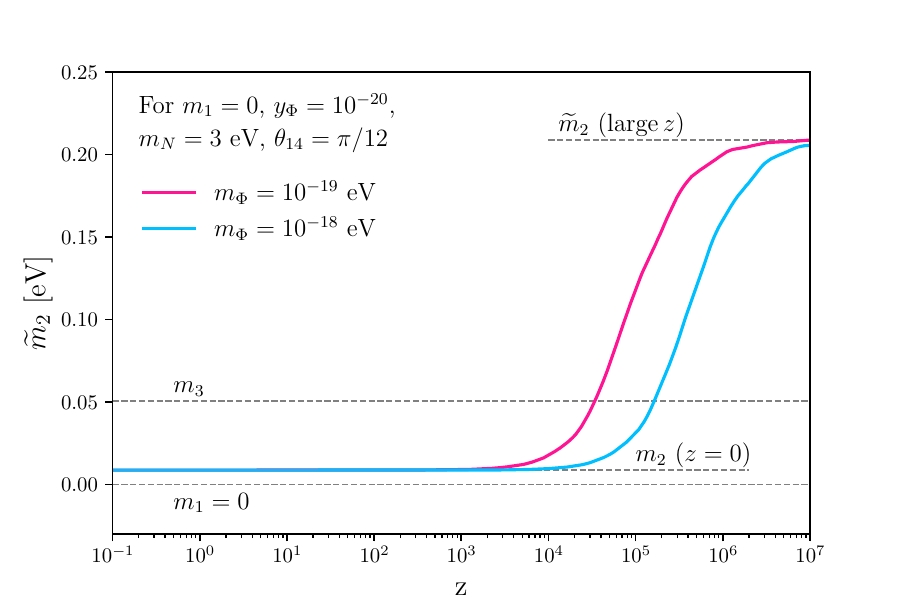}
    \caption{Evolution of the effective mass as a function of redshift in the scenario where the ULDM couples only with $\nu_2$. In this case, $\Tilde{m}_2$ redshifts and can become larger than $m_3$ at large redshifts. The plot assumes normal mass ordering and the lightest neutrino mass $m_1=0$. }
    \label{fig:m2redshift}
\end{figure}

As a demonstration, let us consider a simple case in which the mixing and the structure of the coupling are such that only $\nu_2$ gets an effective mass due to the mixing with $N$. Let us also assume that at present times the mass-ordering is the so-called \textit{normal} and the lightest neutrino, $\nu_1$, is massless, $m_{1} = 0$. Then, according to oscillation data~\cite{deSalas:2020pgw}, one can constrain neutrino masses to be $m_2 (z = 0) = \sqrt{\Delta m^2 _{21}} = 8.7$~meV and $m_3 (z = 0) = \sqrt{\Delta m^2_{31}} = 50$~meV. However, due to the evolution of the dark-matter energy density, $\Tilde{m}_{2}$ could have been larger than $\Tilde{m}_{3}$ in the early Universe, as shown in Fig.\,\ref{fig:m2redshift} for a set of illustrative values. As long as the limit on the sum of neutrino masses is not violated by such a transition, this kind of scenario is extremely difficult to constrain cosmologically. For the example shown in Fig.\,\ref{fig:m2redshift}, we see that $\sum \Tilde{m}_i$ remains virtually unchanged for redshifts up to $z\sim \mathcal{O}(10^3)$, and hence this is consistent with cosmological limits. This can possibly be tested by using neutrino free-streaming arguments~\cite{Taule:2022jrz}, but a dedicated study is beyond of the scope of this work. One should also notice that, in this particular case, the ordering of neutrino masses changes but it does not correspond to the so-called \textit{inverted} ordering, for which $\nu_3$ is the lightest mass eigenstate.

Time and redshift variation in neutrino mass can also show up in the $\CNB$, which consists of neutrinos free-streaming since the time of neutrino decoupling from the thermal plasma. Decoupling occurred when the weak interactions of the SM neutrinos became less frequent compared to the expansion rate of the Universe at a photon temperature of approximately $T_\gamma \simeq 1~\MeV$. Since decoupling, these neutrinos have gradually cooled down due to the expansion of the Universe. At present, they are expected to exhibit a Fermi-Dirac distribution with a temperature of about $T_\nu \simeq 1.95\,{\rm K}$. This temperature is slightly lower than the photon temperature due to electron-positron annihilation at around $0.5\,{\rm MeV}$ that heated up the photon bath approximately by a factor of $(11/4)^{1/3}$.
In this standard scenario, the number density of these neutrinos can be estimated to be 
\begin{equation}
    n_\nu = \frac{3}{4}\frac{\zeta(3)}{\pi^2}gT_\nu^3\,,
\end{equation}
where $g$ is the number of degrees of freedom. Consequently, the expected number density of the $\CNB$~is  $112\,{\rm cm}^{-3}$ per flavor. 

Due to mixing with sterile neutrinos, neutrinos from the $\CNB$ will pick up a time-redshift-dependent contribution to the effective mass. As a result, any experiment trying to detect the $\CNB$ through the effect of a non-zero neutrino mass would be sensitive to this signal. Currently, one of the most promising avenues in the detection of the $\CNB$ is provided by the PTOLEMY collaboration. In the next section, we will study how a PTOLEMY-like experiment can set very competitive constraints on the coupling and masses of ULDM.

\section{Capture rate in a PTOLEMY-like experiment}
\label{sec:PTOLEMY}
PTOLEMY is one of the most realistic proposals for the detection of the $\CNB$. It considers the production of an observable electron after the neutrino capture by a neutron, $\nu_e + n \to p^+ + e^-$. To make the capture experimentally viable, PTOLEMY intends to use tritium as a target due to its large lifetime and its $Q_\beta$ value, of about 12.3 years and $18.591$ keV, respectively~\cite{Cocco:2007za}.
The capture rate of the $\CNB$~on a target nuclei is generally given by~\cite{Long:2014zva}
\begin{align}\label{eq:CR_ptol}
    \Gamma_{\CNB}= N_T\,\overline{\sigma}\,\sum_{i=1}^3 \left[n(\nu_{i,+1})\,{\cal A}_i(+1)+ n(\nu_{i,-1})\,{\cal A}_i(-1)\right]\,,
\end{align}
where $N_T$ are the number of targets, $n(\nu_{i,\pm 1})$ represent the $\CNB$~number densities for each helical state, and ${\cal A}(h)$ are spin-dependent factors that arise due to the mismatch between helicity and chirality,
\begin{align}
    {\cal A}_i(h)\equiv 1-h v_i,
\end{align}
being $v_i=|\vec{p}|/\sqrt{|\vec{p}|^2+m_i^2}$ the average neutrino velocity, $h$ the helicity.
The nucleus-dependent factor $\overline{\sigma}$ in the capture rate is the spin-averaged cross-section. Assuming tritium as the target, we have that 
\begin{align}
    \overline{\sigma} \approx 3.8\times 10^{-45}{\rm\ cm^2}.
\end{align}
Although the number of events is expected to be large, $\sim 8~{\rm yr^{-1}}$ for 100 g of tritium~\cite{Long:2014zva}, the main challenge for the experimental confirmation of the $\CNB$~is the energy resolution~\cite{Cheipesh:2021fmg,PTOLEMY:2022ldz}.
Since tritium is a $\beta$ emitter, it is necessary to have an excellent energy resolution to distinguish the $\CNB$-produced electrons from those from $\beta$ decay.
PTOLEMY is expected to have a resolution of $\Delta=50-150$ meV, with $\Delta = \sqrt{8 \ln 2} \sigma$ the full width at half maximum (FWHM) of the Gaussian resolution~\cite{PTOLEMY:2019hkd,PTOLEMY:2022ldz}.
The spectra of $\CNB$ and $\beta$ decay events, our signal and background respectively, are computed by including a Gaussian resolution function~\cite{Long:2014zva}
\begin{subequations}
    \begin{align}
        \frac{d \Gamma_{\rm C\nu B}}{dE_e} &= \sum_i\frac{1}{\sqrt{2\pi}\sigma} \int  dE_e^\prime \,\Gamma_{\CNB}^i\, \delta(E_e^\prime - E_{\rm end} - 2m_{i}) \exp\left[-\frac{(E_e^\prime - E_e)^2}{2\sigma}\right]\\
        \frac{d \Gamma_{\beta}}{dE_e} &= \sum_i\frac{1}{\sqrt{2\pi}\sigma} \int dE_e^\prime \, \frac{d \Gamma_\beta}{dE_e} (E_e^\prime) \exp\left[-\frac{(E_e^\prime - E_e)^2}{2\sigma}\right]
    \end{align}
\end{subequations}
where $d \Gamma_\beta/dE_e$ is the $\beta$ spectrum from tritium decay, and $E_{\rm end}$ is the beta decay endpoint energy,
\begin{align}
    E_{\rm end} = K_{\rm end} + m_e,
\end{align}
with $m_e$ the electron's mass and 
\begin{align}
    K_{\rm end} = \frac{(m_{\rm ^3H} - m_e) - (m_{\rm ^3H_e}+m_i)^2}{2m_{\rm ^3H}},
\end{align}
the electron endpoint's kinetic energy.
In what follows, we use the $\beta$ decay spectrum given in Ref.~\cite{Ludl:2016ane}.

\begin{figure}[!t]
    \includegraphics[width=\textwidth]{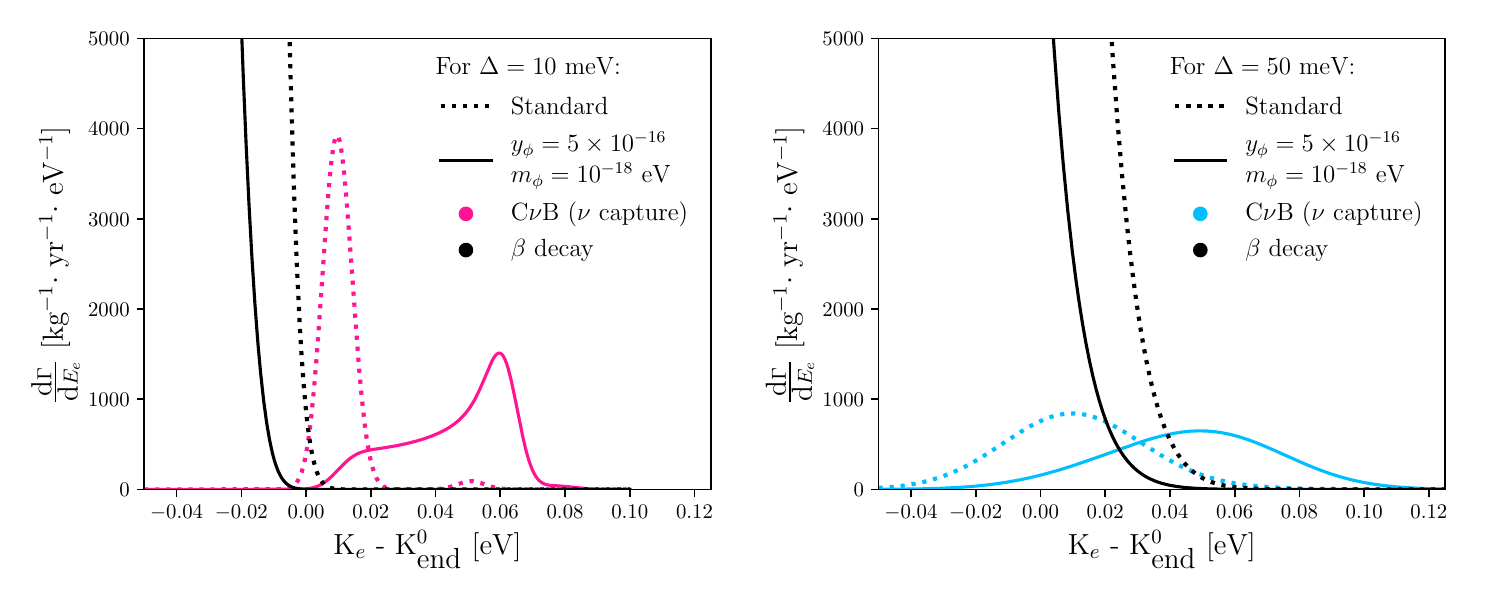}~~
    \caption{Differential electron spectrum from beta decay of tritium and from neutrino capture on tritium. The left and right panels correspond to energy resolutions of $\Delta = 10$ meV and $\Delta =$ 50 meV respectively. Dashed lines represent the standard predictions and the solid lines are the predictions for a coupling $y_\Phi = 5\times 10 ^{-16}$ and an ultralight scalar mass $m_\Phi = 10^{-18}$ eV.  }
    \label{fig:spectra}
\end{figure}
In our scenario of interest, where the neutrino masses vary over time due to the interaction between the $\CNB$~and the ULDM, we anticipate modifications in the spectra measured by a PTOLEMY-like experiment due to these novel interactions.
The primary changes induced by neutrino-ULDM interactions involve alterations in the $\CNB$~capture rate, which is directly influenced by neutrino mass values, and shifts in the spectrum's peak position.
However, it is worth noting that the $\beta$ background is also influenced by mass variation, as the endpoint energies directly correlate with neutrino masses.
In Fig.\,\ref{fig:spectra}, we display spectra for both $\CNB$~and $\beta$ decay electrons under two energy resolution assumptions: $\Delta = 10$ meV (left) and $\Delta = 50$ meV (right). Dashed lines represent the standard scenario, while full lines indicate the inclusion of neutrino-dark matter (DM) interactions with $y_\Phi=5\times 10^{-16}$ for ultralight DM with $m_\Phi = 10^{-18}$ eV. 
The spectra are plotted against electron kinetic energy minus the value of the endpoint kinetic energy assuming massless neutrinos $K_{\rm end}^0$.
Both signal and background show significant modifications due to novel interactions in both panels. Remarkably, under the optimistic energy resolution of $\Delta = 10$ meV, the $\CNB$~spectra widen, with its peak shifting away from the neutrino mass value. Additionally, the $\beta$ decay spectrum shifts away from the $\CNB$~events, leading to a clearer distinction between signal and background.
For the more realistic energy resolution of $\Delta = 50$ meV, observed shifts remain, albeit with smoother spectra. Consequently, these spectra hint at PTOLEMY's potential to test the mass-varying scenario.

To assess PTOLEMY's capability in detecting neutrino-ULDM interactions, we define the region of interest (ROI) as electron energies within the range of $K_e - K_{\rm end}^0 = [0, 0.2]$ eV. This range encompasses the broadening of the $\CNB$~spectra and potential alterations to the $\beta$ decay background.
Events are computed by integrating over the electron's energy in both $\CNB$ and $\beta$ spectra using a bin width of $\Delta$,
\begin{align}
    N_i = \int_{E_e^i}^{E_e^i+\Delta} d E_e\left\{ \frac{d \Gamma_{\rm C\nu B}}{dE_e} + \frac{d \Gamma_{\beta}}{dE_e} \right\}.
\end{align}
Thus, the number of bins is determined by the energy resolution. We consider the following Poissonian test statistics for our analysis,
\begin{align}
    \chi^2 = 2 \sum_{i={\rm bins}} N_i(y_\Phi,m_\Phi) - N_i^{\rm SM} +  N_i^{\rm SM} \ln\left(\frac{N_i(y_\Phi,m_\Phi)}{N_i^{\rm SM}}\right),
\end{align}
being $N_i^{\rm SM}$ and $N_i(y_\Phi,m_\Phi)$ the number of events for the standard case and including neutrino-DM interactions, respectively.
As for the experimental exposure, we assume a setup with 100 g of tritium with an observation time of 10 years. Note that we have not accounted for a local overdensity of neutrinos due to clustering~\cite{Singh:2002de, Ringwald:2004np,Arvanitaki:2022oby}. Such an effect would further enhance the significance of the signal over the background, boosting the sensitivity projections here discussed. However, a dedicated computation of such overdensity for ultralight dark matter would be needed. For simplicity, we do not consider this effect.

\begin{figure}[!t]
    \includegraphics[width=0.7\linewidth]{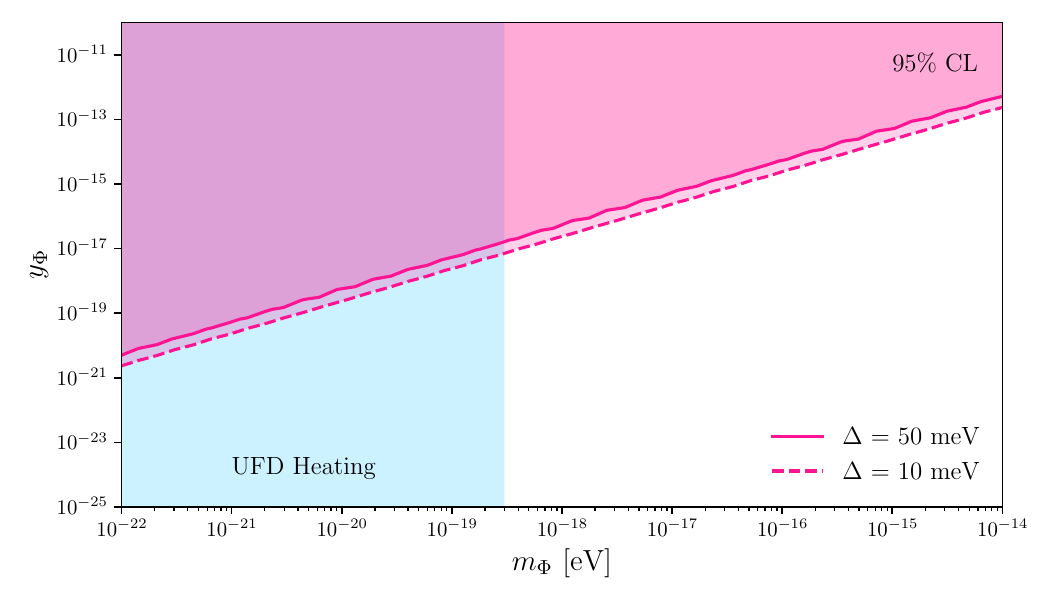}
    \caption{Sensitivity of a PTOLEMY-like experiment to neutrino-ULDM interactions that produce a time-dependent neutrino mass on the coupling $y_\Phi$ vs $m_\Phi$ plane. We present the capabilities to discriminate the additional interaction for two different energy resolutions of $\Delta = 50$ meV (full contour) and $\Delta = 10$ meV (dashed contour) at the $95\%$ CL. We also show the region excluded from ultra-faint white dwarfs heating in the blue regions~\cite{Dalal:2022rmp}.}
    \label{fig:chisqr}
\end{figure}
In Fig.\,\ref{fig:chisqr}, we illustrate PTOLEMY's sensitivity to the parameter space spanned by $y_\Phi$ and $m_\Phi$, assuming the same energy resolutions as previously specified, namely $\Delta = 10$ meV (dashed line) and $\Delta = 50$ meV (full line), at the $95\%$ CL. 
As expected, the optimistic value of $\Delta = 10$ meV offers better sensitivity for our scenario compared to the more realistic energy resolution of $\Delta = 50$ meV. 
Although the $\beta$ background significantly contributes to the event count in our ROI at higher energy resolutions, sensitivity to the mass-varying effect improves with better discrimination of the $\CNB$ signal from the $\beta$ background. 
This is due to the significant dependence of the recoil electron spectra on neutrino masses as previously observed in Fig.\,\ref{fig:spectra}, should neutrino-ULDM interactions exist.
However, we observe that PTOLEMY's capability for setting constraints is only mildly dependent on the energy resolution. 
As a comparison, we also show the constraints obtained from heating ultrafaint dwarf galaxies~\cite{Dalal:2022rmp}. Note that these are stronger than the ones from dwarf galaxies, $m_\Phi > 2.2 \times 10 ^{-22}$ eV at 95\% C.L.~\cite{Zimmermann:2024xvd}. We find that could set more stringent constraints on neutrino-ULDM coupling in part of the parameter space. Model-dependent constraints can also arise from the condition that the ULDM does not redshift like radiation and wash away satellite galaxies of the Milky Way~\cite{Dev:2022bae}. These constraints are competitive with the ones we obtain from PTOLEMY.

\section{Conclusion}
\label{sec:conc}
A hypothetical interaction between neutrinos and an ultralight dark matter scalar field would result in a plethora of experimental signatures. It is well known that a direct coupling of neutrinos with ultralight dark matter can result in the neutrino mass redshifting in past, which can be in tension with the limits on the sum of the neutrino mass derived from the cosmic microwave background. This can be evaded through a sterile neutrino portal, where the dark matter couples directly with sterile neutrinos, and the active neutrinos feel this coupling only through mixing. Such an interplay between ultralight dark matter and active neutrinos can affect the evolution and detection of the cosmic neutrino background. In this work, we have focused on direct and indirect probes of such interacting dark matter with the cosmic neutrino background. 

On the one hand, active neutrinos would get a redshift-dependent effective mass due to their interaction with the ultralight scalar field through mixing with sterile neutrinos. The redshift evolution of the energy density of dark matter would get imprinted in observables related to neutrino masses. In particular, specific coupling structures and values of the active-sterile mixing could lead to a change in the mass ordering of neutrinos between the present time and the early Universe. The phenomenology related to this family of models could also address a discrepancy between indirect -- cosmological -- determinations of the absolute neutrino-mass scale and direct ones -- such as beta decay.

On the other hand, the interaction between neutrinos and dark matter could also manifest in experiments aiming to directly measure the cosmic neutrino background. Specifically, a PTOLEMY-like experiment would observe a distortion in the spectrum from neutrino capture, which could be used to constrain the mass of the dark matter candidate and its coupling to neutrinos. We have presented the projected limits of such an experiment and highlighted the importance of the energy resolution to separate the signal from neutrino capture and beta decay of the target isotope.

The phenomenology and experimental probes here proposed are not unique to the scenario of ultralight scalar dark matter coupling to active neutrinos via mixing with a sterile neutrino. For instance, other scenarios in which neutrinos acquire a redshift-dependent neutrino mass include couplings to dark energy~\cite{Fardon:2003eh,Brookfield:2005bz}, phase transitions~\cite{Lorenz:2018fzb} and topological defects~\cite{Dvali:2021uvk}. In this case, one has to ensure that the specific scenario is consistent with all cosmological probes, i.e. cosmic microwave background, Big Bang nucleosynthesis, and the observed large- and small-scale structures. Similarly, one can consider scenarios in which neutrino mass shows a similar time-modulation at the present time, or other time-dependent phenomena, for instance, in the case of Lorentz Invariance Violation~\cite{Kostelecky:2004hg,Diaz:2009qk,Diaz:2014yva}. Finally, other ultralight dark matter candidates include vector or tensor fields (e.g. \cite{LopezNacir:2018epg,Marzola:2017lbt}). In each of those cases, a coupling to neutrinos would also modify the evolution of the cosmic microwave background and its detection prospects. In each of the cases, the signatures differ from those of coupling to scalars and hence are not discussed in this article. Nonetheless, the methodology and part of the discussions here presented would also apply to them.

\section*{Acknowledgements}

We would like to thank Tim Herbermann, Joachim Kopp, and Alexei Smirnov for useful discussions.
PMM acknowledges support from the Carlsberg Foundation (CF18-0183). 
The work of YFPG has been funded by the UK Science and Technology Facilities Council (STFC) under grant ST/T001011/1. 
This project has received funding/support from the European Union’s Horizon 2020 research and innovation programme under the Marie Sk\l{}odowska-Curie grant agreement No 860881-HIDDeN.
This work has made use of the Hamilton HPC Service of Durham University.

\bibliography{references}

\end{document}